\documentclass[twocolumn,superscriptaddress,amssymb,amsmath,nobibnotes,aps,prlgroupedaddress]{revtex4}  

\usepackage{graphicx}  
\usepackage{dcolumn}   
\usepackage{bm}        
\usepackage{graphics,epsfig}

\usepackage[utf8]{inputenc}
\usepackage[english]{babel}
\makeatletter
\usepackage{mathtools}
\usepackage{xcolor}

\usepackage{wasysym} 
\usepackage{float} 
\usepackage{setspace} 
\usepackage{indentfirst} 
\usepackage{cancel} 
\usepackage{hyperref} 
\hypersetup{
            colorlinks=true,
            linkcolor=black,
            urlcolor=blue,
            }

\usepackage{esint} 
\usepackage{tikz} 
\usetikzlibrary{babel} 
\usepackage{tcolorbox} 

\DeclareMathOperator{\tr}{Tr}

\newcommand{\of}[1]{\left(#1\right)}
\newcommand{\off}[1]{\left[#1\right]}

\newcommand{\e}[1]{e^{#1}}

\begin{document}


\title{Rényi entropies of the massless Dirac field on the torus}

\author{David Blanco}
\email{dblanco@df.uba.ar}
\affiliation{Departamento de F\'{\i}sica, FCEN, Universidad de Buenos Aires and IFIBA-CONICET\\
1428 Buenos Aires, Argentina}
\author{Tomás Ferreira Chase}
\affiliation{Departamento de F\'{\i}sica, FCEN, Universidad de Buenos Aires\\
1428 Buenos Aires, Argentina}
\author{Juan Laurnagaray}
\affiliation{Departamento de F\'{\i}sica, FCEN, Universidad de Buenos Aires\\
1428 Buenos Aires, Argentina}
\author{Guillem P\'erez-Nadal}
\email{guillem@df.uba.ar}
\affiliation{Departamento de F\'{\i}sica, FCEN, Universidad de Buenos Aires and IFIBA-CONICET\\
1428 Buenos Aires, Argentina}

\date{\today}

\begin{abstract}
We compute the Rényi entropies of the massless Dirac field on the Euclidean torus (the Lorentzian cylinder at non-zero temperature) for arbitrary spatial regions. 
We do it by the resolvent method, i.e., we express the entropies in terms of the resolvent of a certain operator and then use the explicit form of that resolvent, which was obtained recently. Our results are different in appearance from those already existing in the literature (obtained via the replica trick), but they agree perfectly, as we show numerically for non-integer order and analytically for integer order. We also compute the Rényi mutual information, and find that, for appropriate choices of the parameters, it is non-positive and non-monotonic. This behavior is expected, but it cannot be seen with the simplest known Rényi entropies in quantum field theory because they are proportional to the entanglement entropy.
\end{abstract}

\maketitle

\section{Introduction}

In the last years, the use of ideas and results coming from quantum information theory has provided deep insights into the properties of quantum field theory (QFT).
Some examples are the proof of the irreversibility of the renormalization group flow in various dimensions \cite{Casini:2006es,Casini:2012ei,Casini:2015woa,Casini:2017vbe}, the formulation of a well-defined version of the Bekenstein bound \cite{Casini:2008cr} and several energy inequalities \cite{Blanco:2013lea,Faulkner:2016mzt,Blanco:2017akw,Balakrishnan:2017bjg}. There have also been many applications to holography (see \cite{VanRaamsdonk:2016exw} for a review) and, more recently, to the black hole information problem \cite{Penington:2019npb,Almheiri:2019psf,Almheiri:2020cfm}.

In this context, one is typically interested in quantifying the degree of mixing of a density matrix $\rho$ obtained by reducing a given global state to a certain spatial region. The usual measure is the von Neumann entropy (which for reduced states is also called entanglement entropy), but a more complete characterization of the degree of mixing of $\rho$ is provided by the Rényi entropies,
\begin{equation}\label{renyi}
    S_n=\frac{1}{1-n}\log\left(\tr\rho^n\right),
\end{equation}
where $n$ is a positive real number called the order of the Rényi entropy. 
The entanglement entropy is just one element of the family, $n=1$. The Rényi entropies are positive and monotonically decreasing with $n$, so they have a limit as $n\to\infty$. This limit is called the min-entropy, and it gives the largest eigenvalue, $p_\text{max}$, of $\rho$, $S_\infty=-\log p_\text{max}$.

In the context of QFT, the Rényi entropies are ultraviolet divergent, so they must be regulated with a short-distance cutoff. However, with the Rényi entropies of multicomponent regions one can construct a finite quantity, the Rényi mutual information (RMI)
\begin{equation}
    I_n(A:B)=S_n(A)+S_n(B)-S_n(A\cup B),
\end{equation}
where $A$ and $B$ denote spatial regions. In recent years, a lot has been learned about Rényi entropies of multicomponent regions for different theories, mainly conformal field theories (CFTs). A partial list of references includes \cite{Calabrese:2009ez,Calabrese:2010he,Headrick:2010zt,Hartman:2013mia,Chen:2013kpa,Coser:2013qda,Headrick:2015gba} in $1+1$ dimensions and \cite{Cardy:2013nua,Agon:2015ftl,Chen:2016mya} in higher dimensions. However, exact results for arbitrary regions and arbitrary $n\in{\mathbb R}^+$ are only available for the massless Dirac field in $1+1$ dimensions. There is a result \cite{Casini:2005rm} on the Euclidean plane (i.e., Minkowski spacetime at zero temperature), from which one can obtain results on the Euclidean cylinders (Minkowski spacetime at non-zero temperature and the Lorentzian cylinder at zero temperature) by a conformal transformation. This result was later generalized to the Euclidean torus (the Lorentzian cylinder at non-zero temperature) in \cite{Herzog:2013py} (see also \cite{Azeyanagi:2007bj}).

The method used in the above references is the replica trick, which is based on the observation that Rényi entropies of integer order $n\ne 1$ are related to the partition function of the theory on a non-trivial manifold. One computes this partition function, thereby obtaining the integer Rényi entropies, and then analytically continues to arbitrary $n$.
In the case of the torus, the last step is not obvious at first sight, because $n$ enters in the integer Rényi entropies as the number of terms of a sum. Nevertheless, the authors of \cite{Herzog:2013py,Azeyanagi:2007bj} managed to extend their results to arbitrary $n$ in the form of an infinite series.

In a recent paper \cite{Blanco:2019cet} (see also \cite{Fries:2019acy}), we computed the entanglement entropy ($n=1$) of the massless Dirac field on the torus by a different method, other than the replica trick, and we obtained a result which is formally different from that of \cite{Herzog:2013py,Azeyanagi:2007bj} (it involves an integral instead of a series) but nevertheless agrees with it.
The starting point of the method is a general relation \cite{Casini:2009sr} (valid for free field theories in Gaussian states) between the entanglement entropy of a region $V$ and the two-point function $G$, restricted to pairs of points in $V$ and viewed as an operator which acts on functions by convolution. This relation can be reexpressed in terms of the resolvent of $G$, so if one manages to compute the resolvent for the theory and state of interest one can obtain the entropy. The advantage of this method is that it yields directly the result for $n=1$, without need of analytically continuing from other values of $n$. In \cite{Blanco:2019xwi} (see also \cite{Fries:2019ozf}) we computed the resolvent for the massless Dirac field on the torus, and in \cite{Blanco:2019cet} we used it to obtain the entanglement entropy.

The purpose of this paper is to extend the above analysis to Rényi entropies of arbitrary order. As in the case of the entanglement entropy, there is a general relation between the Rényi entropies and the correlator $G$ for free field theories in Gaussian states. We will rewrite this relation in terms of the resolvent of $G$, and then use the resolvent computed in \cite{Blanco:2019xwi} to obtain the Rényi entropies of the massless Dirac field on the torus. Our results are valid for arbitrary $n\in{\mathbb R}^+$ and arbitrary spatial regions. Again, they are formally different from those of \cite{Herzog:2013py,Azeyanagi:2007bj}, but we will show that they agree, numerically for non-integer $n$ and analytically for integer $n$. We will also show that our general formula reduces to our previous result for the entanglement entropy in the case $n=1$, and study the limits where one of the periods of the torus is sent to infinity so that it becomes a cylinder. Finally, we will use our results for the Rényi entropies to compute the RMI in simple cases.

As is well-known, the entanglement entropy is subadditive and strongly subadditive; in other words, the mutual information is positive and monotonic (i.e., it grows with the size of the subsystems). These properties are in general not shared by the RMI with $n\ne 1$. For example, for two qubits $A$ and $B$ in the separable state
\begin{equation}
    \rho_{AB}=\sum_{i,j=0}^1 p_{ij}|ij\rangle\langle ij|
\end{equation}
with $p_{00}=p_{01}=p_{10}=1/3$ and $p_{11}=0$, it is straightforward to see that $I_\infty(A:B)=\log(3/4)<0$. Moreover, since the Rényi entropies of the empty subsystem vanish, we have $I_\infty(A:\varnothing)=0>I_\infty(A:B)$, thus breaking monotonicity. As far as we know, no example like this of a non-positive and non-monotonic RMI has been reported in the context of QFT. The simplest known Rényi entropies (those of the massless Dirac field on the plane, or their conformal transformations to the cylinders) do not provide such an example, because they are proportional to the entanglement entropy with a constant coefficient (independent of the spatial region) and hence they are subadditive and strongly subadditive. One might naively wonder whether this is a general property of QFT. We will rule out this possibility by showing that, for appropriate choices of the parameters, our RMIs are non-positive and non-monotonic.

The paper is organized as follows. In section \ref{sect:2} we present the general relation between the Rényi entropies of Dirac fields in Gaussian states and the correlator $G$, and rewrite it in terms of the resolvent of $G$. In section \ref{sect:3} we particularize to the massless field on the torus, and use the explicit form of the resolvent to compute the Rényi entropies. In section \ref{sect:4} we explore some particular cases, namely the limits where the torus becomes a cylinder, the case $n=1$ and the other integer values of $n$. In section \ref{sect:5} we compute the RMI in simple cases and study its positivity and monotonicity properties. And we close in section \ref{sect:6} with a discussion of our results. The paper also contains two appendices: in appendix \ref{appendixB} we give the relation between the Weierstrass and the Jacobi theta functions, which is useful for comparison with previous literature, and in appendix \ref{appendixC} we discuss the properties of the Weierstrass elliptic function, which we use to study the positivity and monotonicity of the RMI.

\section{Rényi entropies of a Dirac field in a Gaussian state}\label{sect:2}

\begin{figure}
    \centering
    \includegraphics[width=0.43\textwidth]{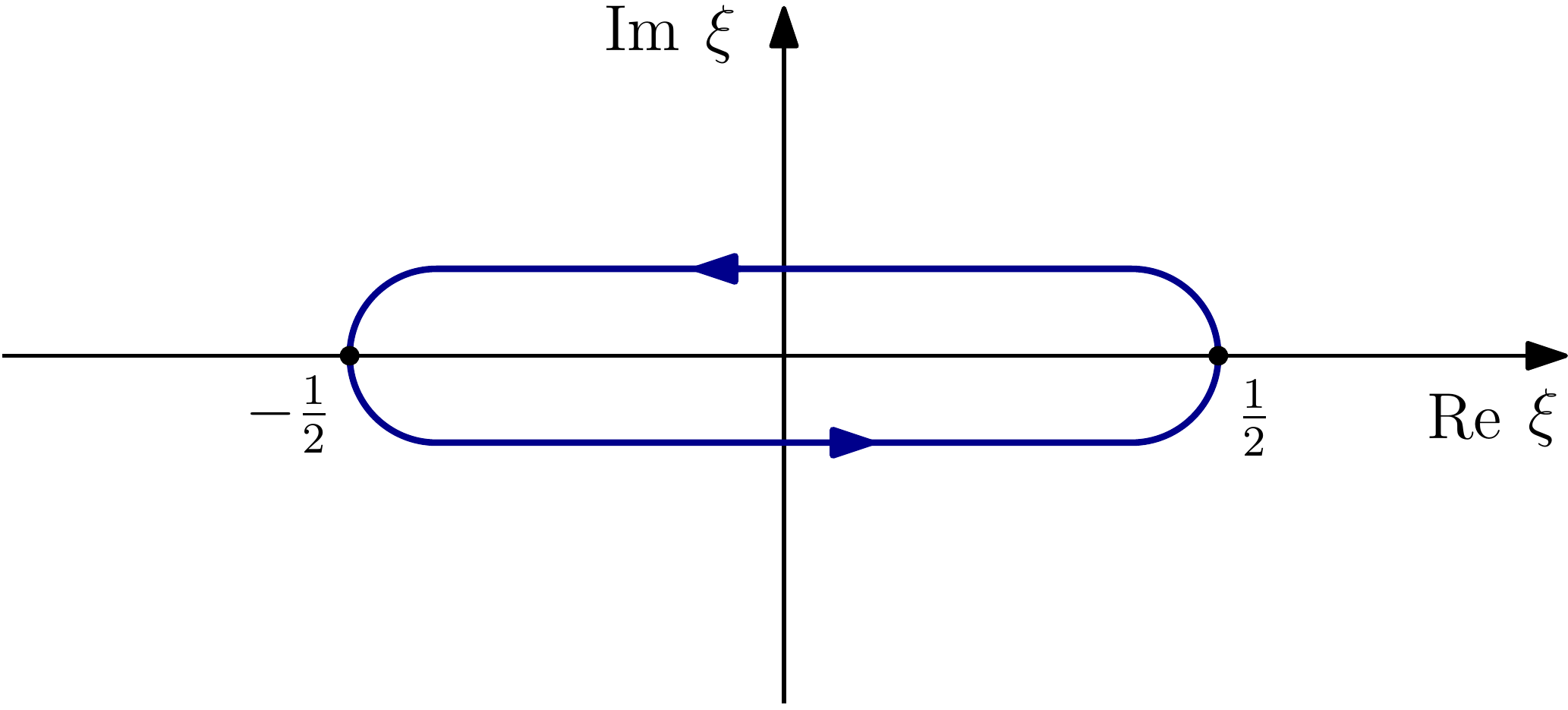}
    \caption{Contour of integration for equation (\ref{renyiR}). The distance between the two horizontal stretches is infinitesimally small. There is no problem at the branch points $\xi=\pm 1/2$ because the integrand is continuous there.}
    \label{fig1}
\end{figure}

Consider a Dirac field $\psi$ in a Gaussian state, i.e., a state in which correlation functions obey Wick's theorem. The reduction of such a state to a spatial region $V$ is known to have the form \cite{peschel2003calculation,Casini:2009sr}
\begin{equation}\label{density}
    \rho=\frac{1}{Z}\exp\left[-\int_V d^dx\, d^dy\,\psi^\dagger(x)K(x,y)\psi(y)\right],
\end{equation}
where $Z$ is a normalization constant, $d$ is the spatial dimension and the kernel $K$ is given in terms of the correlator $G_{ij}(x,y)=\langle\psi_i(x)\psi_{j}^\dagger(y)\rangle$ ($x,y\in V$) by the equation
\begin{equation}\label{modG}
    K=-\log\left(G^{-1}-1\right).
\end{equation}
Here, both $K$ and $G$ are viewed as operators acting on vector-valued functions on $V$ (any matrix-valued function $M$ of two variables naturally defines such an operator via the equation $(M v)(x)=\int dy\,M(x,y)v(y)$). Note that $G$ is positive and, due to the canonical anticommutation relations, so is $1-G$. This implies that the spectrum of $G$ is contained in the interval $(0,1)$, which in turn guarantees that $K$ is Hermitian.
Substituting equations (\ref{density}) and (\ref{modG}) into (\ref{renyi}) one obtains a formula for the Rényi entropies of $\rho$ \cite{Casini:2009sr},
\begin{equation}
    S_n=\frac 1{1 - n} \tr\log\left[G^n + \of{1 - G}^n\right]\,.\label{renyiC}
\end{equation}
Our first purpose is to rewrite this equation in terms of the resolvent of $G$,
\begin{equation}\label{resolventdef}
    R(\xi)=\frac{1}{\xi-(1/2-G)}\,.
\end{equation}
For that purpose, recall Cauchy's integral formula: if $f$ is an analytic function, then
\begin{equation}
    f(1/2-z)=\frac{1}{2\pi i}\ointctrclockwise d\xi\,\frac{f(\xi)}{\xi-(1/2-z)}
\end{equation}
for any contour enclosing the point $1/2-z$. Defining
\begin{equation}
    f_n(\xi)=\frac{1}{1-n}\log\left[(1/2-\xi)^n+(1/2+\xi)^n\right],
\end{equation}
equation (\ref{renyiC}) takes the form $S_n=\tr f_n(1/2-G)$, so, by Cauchy's integral formula, we can write
\begin{equation}\label{renyiR}
    S_n=\frac{1}{2\pi i}\tr\ointctrclockwise d\xi\,R(\xi)f_n(\xi)
\end{equation}
for any contour enclosing the spectrum of $1/2-G$ and in whose interior $f_n$ is analytic. For generic $n$, this function has cuts on the half-lines $(-\infty,-1/2)$ and $(1/2,\infty)$, and also at the points where the argument of the logarithm is real negative; everywhere else, $f_n$ is analytic. Taking into account that the spectrum of $1/2-G$ is contained in the interval $(-1/2,1/2)$ and that the argument of the logarithm does not become real negative anywhere near it (because, for $\xi$ in this interval, $(1/2-\xi)^n+(1/2+\xi)^n>\text{min}\{1,2^{1-n}\}$), a good contour is that of figure \ref{fig1}.
This completes our task of writing the Rényi entropies in terms of the resolvent of $G$. 

The analysis of this section was quite general: it applies to Dirac fields of any mass in any number of dimensions, and to any Gaussian state. In the next sections we will be more specific.

\section{Rényi entropies of the massless field on the torus}\label{sect:3}

\subsection{Massless field on the torus}

Let us now specialize to a massless field in $1+1$ dimensions. In this case $\psi=(\psi_+,\psi_-)$ is a two-component spinor, and the two components (chiralities) decouple from each other. We take the spatial manifold to be a circle of length $L$, so $\psi$ satisfies the boundary condition
\begin{equation}
    \psi(x+L)=(-1)^\nu\psi(x)
\end{equation}
with $\nu\in\{0,1\}$. The periodic case $\nu=0$ is also called a Ramond boundary condition, whereas the antiperiodic case $\nu=1$ is called a Neveu-Schwarz boundary condition. We also choose a particular Gaussian state: a thermal state with inverse temperature $\beta$. The correlator in this case has the form $G={\text{diag}}(G_+,G_-)$ with
\begin{equation}\label{correlator}
    G_\pm(x,y)=\pm\frac{1}{2\pi i}\frac{\sigma_\nu(x-y)}{\sigma(x-y\mp i\epsilon)}.
\end{equation}
In this equation, $\sigma$ is the Weierstrass sigma function associated with the complex torus of periods $L$ and $i\beta$,
\begin{equation}\label{sigma}
    \sigma(z)=z\prod_{\lambda\ne 0}\left[\left(1+\frac{z}{\lambda}\right)e^{-\frac{z}{\lambda}+\frac{1}{2}\left(\frac{z}{\lambda}\right)^2}\right],
\end{equation}
where $\lambda$ runs over the lattice $\Lambda=\{m L+n i\beta,m,n\in{\mathbb Z}\}$. A related function is the Weierstrass zeta function,
\begin{equation}
    \zeta=\frac{\sigma'}{\sigma},
\end{equation}
and both functions enter in the definition of $\sigma_\nu$,
\begin{equation}
    \sigma_\nu(z)=e^{-[\zeta(L/2)+\nu\zeta(i\beta/2)]z}\frac{\sigma(z+L/2+\nu i\beta/2)}{\sigma(L/2+\nu i\beta/2)}.
\end{equation}
Equation (\ref{correlator}) can be rewritten in a perhaps more familiar form by using the relation between the Weierstrass functions and the Jacobi theta functions, which we give in appendix \ref{appendixB}.

In order to gain familiarity with the Weierstrass functions, let us briefly discuss some of their properties (see \cite{chandrasekharan2012elliptic,Pastras:2017wot} for more details). As is clear from its definition, $\sigma$ is analytic throughout the complex plane (the infinite product in (\ref{sigma}) has no problems of convergence because the factors tend quickly to $1$ as $|\lambda|\to\infty$). It has a zero at the origin, where it has unit derivative, $\sigma'(0)=1$, 
and the remaining zeros are the points congruent to the origin, i.e., which differ from it by an element of the lattice $\Lambda$.
It is odd, $\sigma(-z)=-\sigma(z)$, and commutes with complex conjugation, $\sigma^*(z)=\sigma(z^*)$, so it is real on the real axis and imaginary on the imaginary axis. Moreover, one can show that it is quasiperiodic,
\begin{equation}\label{sigmap}
    \sigma(z+P_i)=-e^{\zeta(P_i/2)(2z+P_i)}\sigma(z),
\end{equation}
where $P_1=L$ and $P_2=i\beta$.
As a consequence of these properties, $\zeta$ is analytic except for poles of unit residue at the points congruent to the origin, it is odd, commutes with complex conjugation and has the quasiperiodicity
\begin{equation}\label{zetap}
    \zeta(z+P_i)=\zeta(z)+2\zeta(P_i/2).
\end{equation}
Integrating $\zeta$ along a rectangle of sides $L$ and $i\beta$, and using the above properties, one obtains a useful relation between the values of $\zeta$ at the half-periods,
\begin{equation}\label{halfperiods}
    i\beta\zeta(L/2)-L\zeta(i\beta/2)=i\pi.
\end{equation}
As for $\sigma_\nu$, it is clearly analytic throughout the complex plane with zeros at the points congruent to $L/2+\nu i\beta/2$, and satisfies $\sigma_\nu(0)=1$. Using the quasiperiodicity of $\sigma$ one finds that it is even, $\sigma_\nu(-z)=\sigma_\nu(z)$, and commutes with complex conjugation, and with the help of (\ref{halfperiods}) one also obtains its quasiperiodicity,
\begin{equation}\label{sigmanup}
    \sigma_\nu(z+P_i)=(-1)^{\nu_i+1}e^{\zeta(P_i/2)(2z+P_i)}\sigma_\nu(z),
\end{equation}
where $\nu_1=\nu$ and $\nu_2=1$. Finally, it is helpful to see what these functions look like when one of the periods is sent to infinity. In this limit, the infinite product in (\ref{sigma}) becomes basically the Euler product formula for the sine and one obtains
\begin{equation}\label{sigmainf}
    \sigma(z)=\frac{P}{\pi}\sin\left(\frac{\pi z}{P}\right)e^{\frac{1}{6}\left(\frac{\pi z}{P}\right)^2},
\end{equation}
where $P$ is the period that stays finite. It follows that
\begin{equation}
    \zeta(z)=\frac{\pi}{P}\left[\cot\left(\frac{\pi z}{P}\right)+\frac{1}{3}\frac{\pi z}{P}\right].
\end{equation}
From the above two equations one obtains the behavior of $\sigma_\nu$, which depends on which of the periods is sent to infinity,
\begin{equation}\label{sigmanuinf}
   \sigma_\nu(z)=\begin{cases}
    e^{-\frac{1}{6}\left(\frac{\pi z}{\beta}\right)^2} &\quad L\to\infty\\
    \cos\left(\delta_{\nu 0}\frac{\pi z}{L}\right)e^{\frac{1}{6}\left(\frac{\pi z}{L}\right)^2} &\quad \beta\to\infty.
    \end{cases}
\end{equation}
Note that the asymmetry between both limits only occurs in the case $\nu=0$, due to the Kronecker delta above. For some plots of the Weierstrass functions, see appendix A of \cite{Blanco:2019cet}.

\subsection{Rényi entropies}

\begin{figure}
    \centering
    \includegraphics[width=0.43\textwidth]{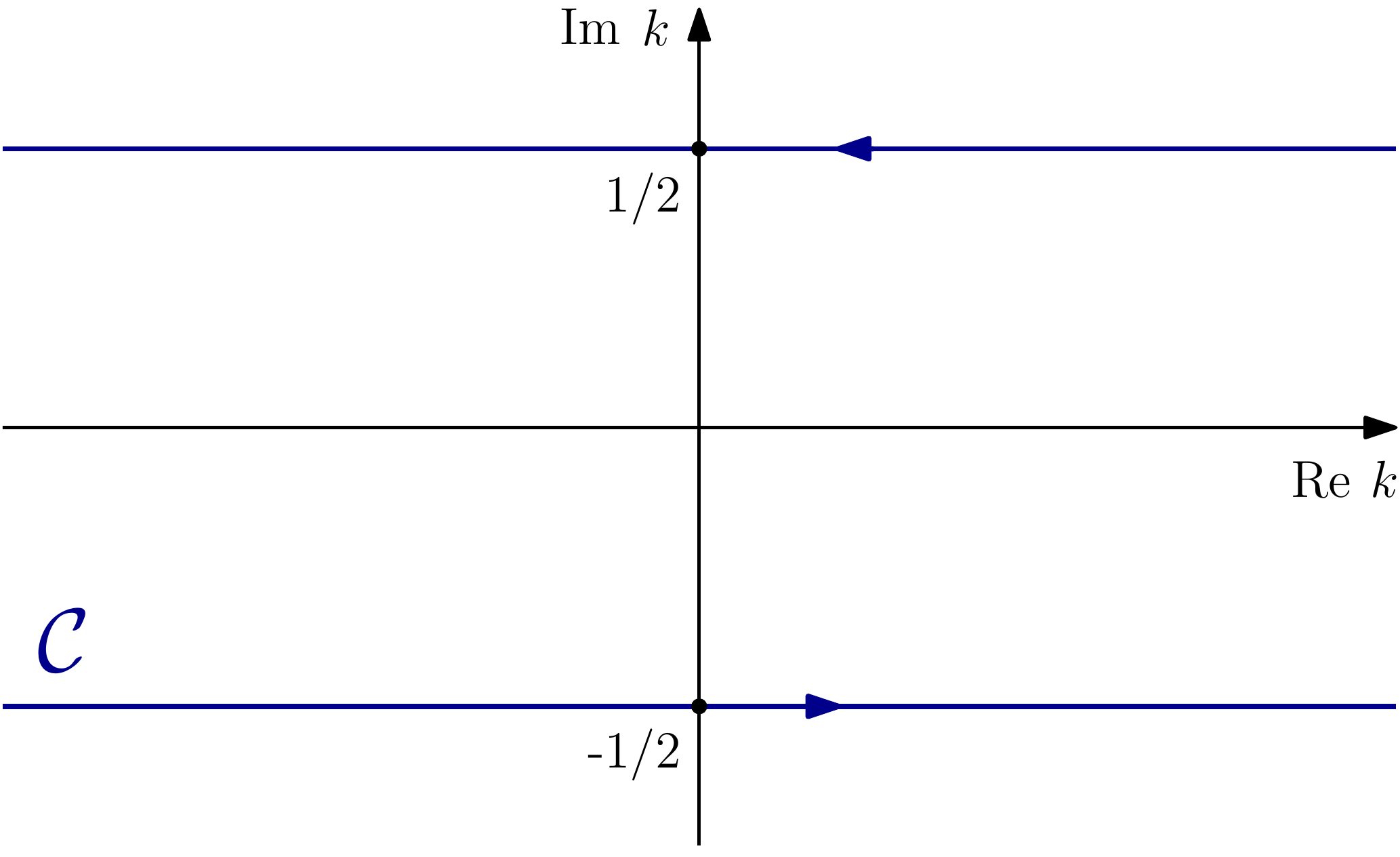}
    \caption{Integration path ${\mathcal C}$, first appearing in equation (\ref{renyik}). Both lines extend to infinity.}
    \label{fig2}
\end{figure}

\begin{figure}
    \centering
    \includegraphics[width=0.4\textwidth]{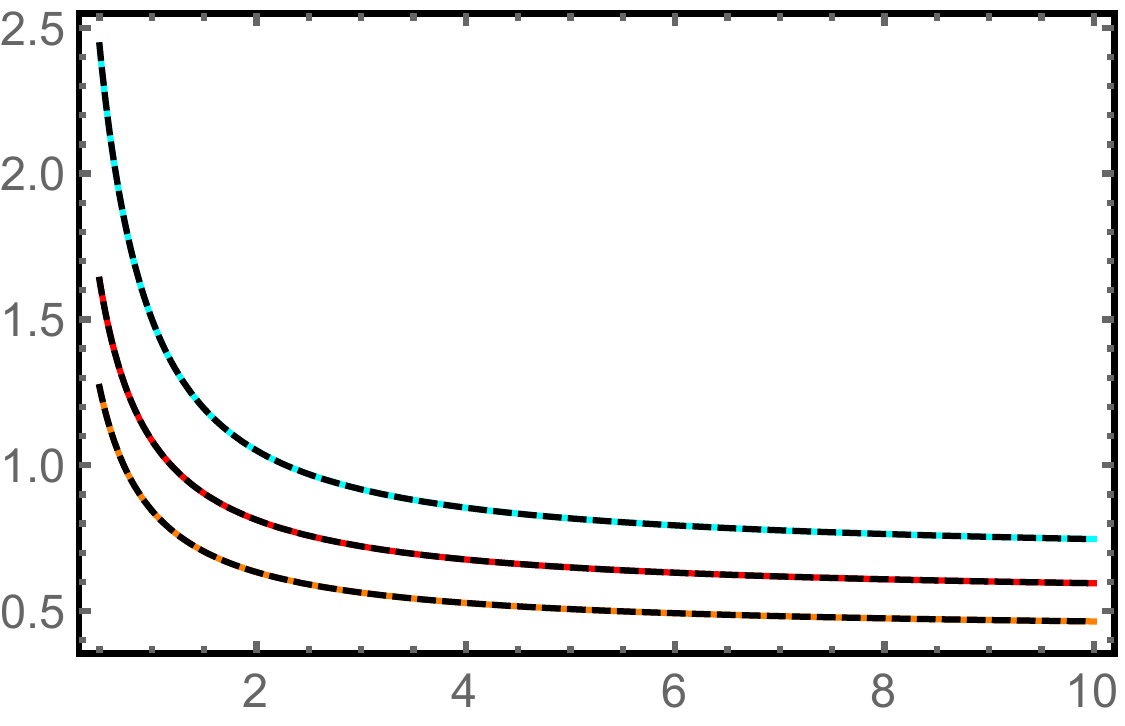}
    \caption{{\bf Rényi entropies as a function of $n$.} We plot the Rényi entropy of one interval of length $\ell$ as a function of $n$ for the values $\ell=1/8$ (bottom), $\ell =1/4$ (middle) and $\ell=3/4$ (top), with $L=1$, $\beta=9/10$, $\epsilon=1/100$ and $\nu=1$. The negative slope of these curves is a general property of the Rényi entropy.}
    \label{fig3}
\end{figure}

\begin{figure}
    \centering
    \includegraphics[width=0.4\textwidth]{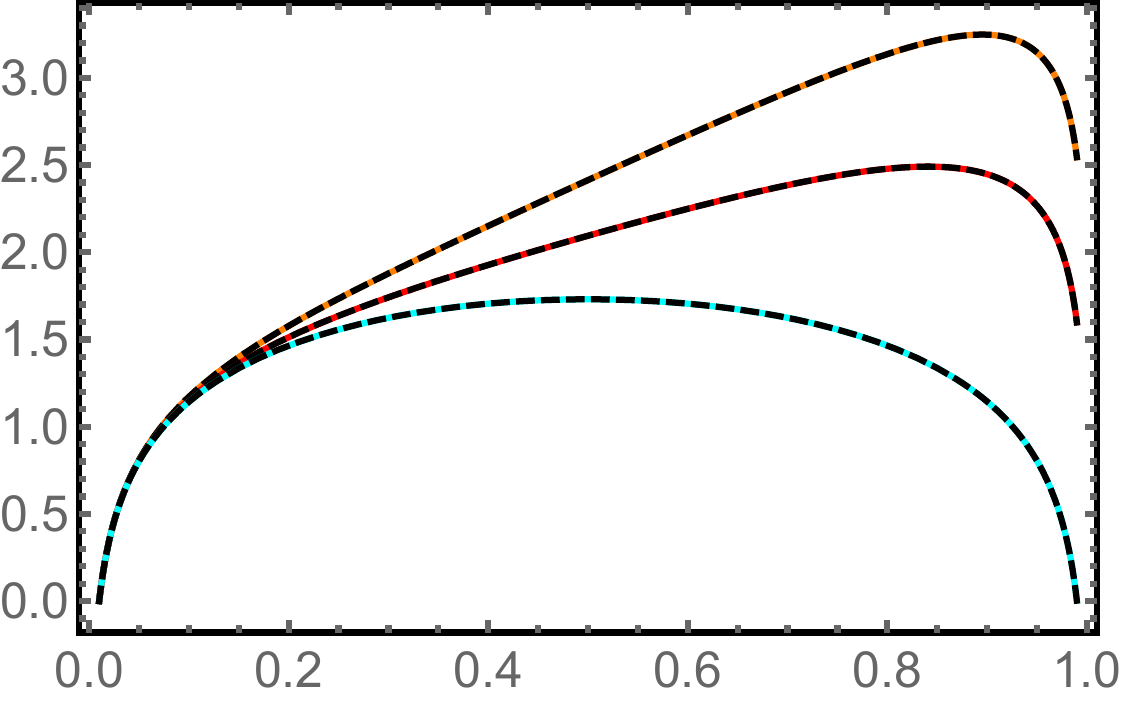}
    \caption{{\bf Neveu-Schwarz boundary conditions.} We plot the $n=1/2$ Rényi entropy for one interval as a function its length $\ell$ for $\beta=6/10$ (top), $\beta =9/10$ (middle) and $\beta=5$ (bottom), with $L=1$, $\epsilon=1/100$ and $\nu=1$. In the limit of zero temperature (bottom curve) the plot is symmetric with respect to the middle point of the interval, as corresponds to a pure global state (the vacuum).}
    \label{fig4}
\end{figure}

\begin{figure}
    \centering
    \includegraphics[width=0.4\textwidth]{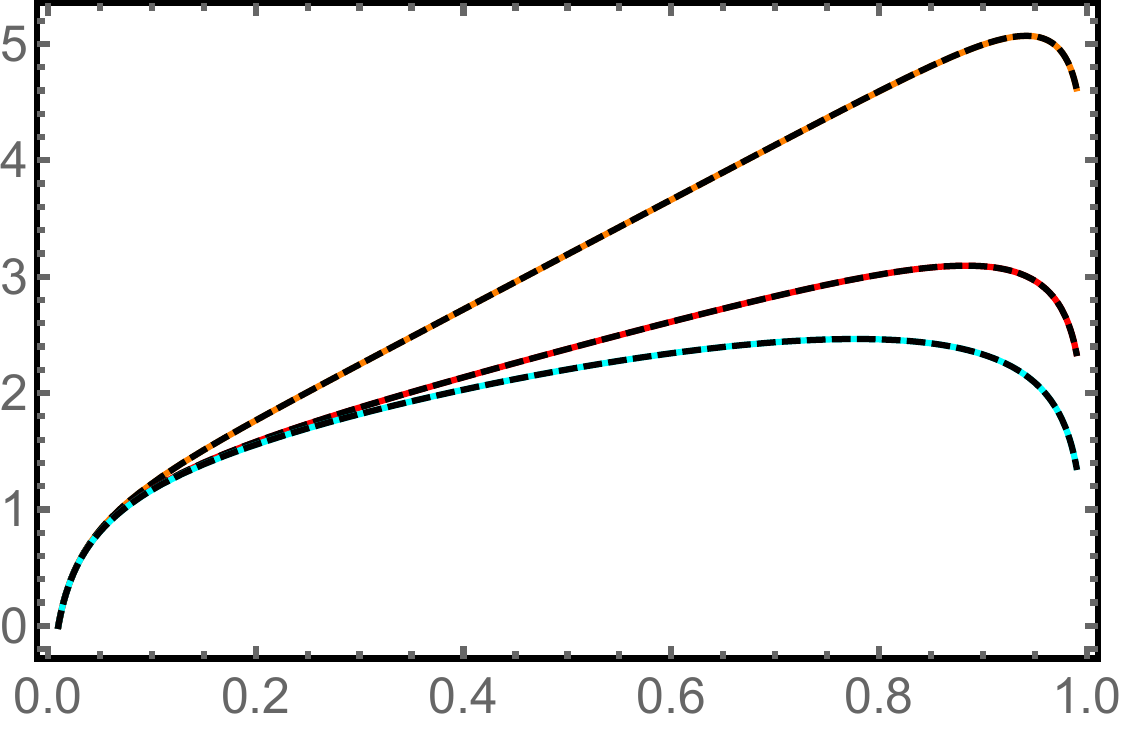}
    \caption{{\bf Ramond boundary conditions.} We plot the $n=1/2$ Rényi entropy for one interval as a function its length $\ell$ for $\beta=1/3$ (top), $\beta =2/3$ (middle) and $\beta=10$ (bottom), with $L=1$, $\epsilon=1/100$ and $\nu=0$. In the limit of zero temperature (bottom curve), the plot is not symmetric with respect to the middle point of the interval. This is due to the degeneracy of the ground state in the Ramond case, which implies that the thermal state remains mixed in the limit of zero temperature.}
    \label{fig5}
\end{figure}

The resolvent of the correlator (\ref{correlator}) was recently computed in \cite{Blanco:2019xwi,Blanco:2019cet,Fries:2019ozf} for an arbitrary subset $V$ of the circle consisting of $N$ disjoint intervals $(a_1,b_1),\dots,(a_N,b_N)$. It has the form $R={\text{diag}}(R_+,R_-)$ with
\begin{eqnarray}
	R_\pm\of{\xi; x, y} &=& \frac 1{\xi^2 - \frac 14} \bigg[\xi \delta\of{x - y}  \label{eq:renyi_reduced:torusresolvent}\\ 
    &\mp& \frac{\e{\mp i k\off{\omega\of x - \omega\of y}}}{2 \pi i \sigma\of{x - y}} \frac{\sigma_\nu\of{x - y \pm i k \ell}}{\sigma_\nu\of{\pm i k \ell}}\bigg]\,,\nonumber
\end{eqnarray}
where $\ell=\sum_{i=1}^{N} (b_i-a_i)$ is the total length of $V$ and
\begin{eqnarray}
    &&k = \frac{1}{2\pi} \log \frac{\xi-\frac{1}{2}}{\xi+\frac{1}{2}}\,,\label{cov}\\
    &&\omega(x) = \sum_{i=1}^{N} \log \left| \frac{\sigma\left( a_i-x\right)}{\sigma\left( b_i-x\right)}\right|.
\end{eqnarray}
In order to obtain the Rényi entropies, we simply have to insert this result into equation (\ref{renyiR}). Since that equation involves a trace, we will ultimately set $x=y$, so it is convenient to expand the second term in the resolvent in powers of $x-y$. This gives
\begin{eqnarray}\label{expansion}
    R_\pm(\xi;x,y)&=&\frac 1{\xi^2 - \frac 14} \bigg\{\xi \delta\of{x - y}\mp\frac{1}{2\pi i(x-y)}\nonumber\\
    &&\!\!\!\!\!\!\!\!\!\!\!\!\!\!\!\!\!\!\!\!\!+\,\frac{1}{2\pi i}\left[ik\omega'(y)-\zeta_\nu(ik\ell)\right]\bigg\}+{\mathcal O}(x-y),
\end{eqnarray}
where we have defined
\begin{eqnarray}
    \zeta_\nu(z) &=& \frac{\sigma_{\nu}'(z)}{\sigma_\nu(z)}\label{zetanu}\\
    &=&\zeta(z+L/2+\nu i\beta/2)-\zeta(L/2)-\nu\zeta(i\beta/2).\nonumber
\end{eqnarray}
Note from the properties of $\sigma_\nu$ (or alternatively from those of $\zeta$) that $\zeta_\nu$ is analytic except for poles at the points congruent to $L/2+\nu i\beta/2$, it is odd, commutes with complex conjugation and has the same quasiperiodicity as $\zeta$. Now, the first two terms in (\ref{expansion}) do not contribute to the Rényi entropy, because the corresponding terms in the integrand of (\ref{renyiR}) are analytic in $\xi$ inside the contour and hence integrate to zero (note that there is no problem at the points $\xi=\pm 1/2$ of the contour, because $f_n$ vanishes there). The remaining two terms in (\ref{expansion}) survive, because $k$ has a cut on the interval $(-1/2,1/2)$, so the Rényi entropy is
\begin{equation}\label{renyiintx}
    S_n=-\frac{1}{2\pi^2}\int_V dx\ointctrclockwise \frac{d\xi}{\xi^2-\frac{1}{4}}\left[ik\omega'(x)-\zeta_\nu(ik\ell)\right]f_n(\xi).
\end{equation}
The integral in $x$ is straightforward,
\begin{equation}\label{intx}
    \int_Vdx\,\omega'(x)=\sum_{i=1}^N\left[\omega(b_i-\epsilon)-\omega(a_i+\epsilon)\right]=2\Xi_\sigma,
\end{equation}
where we define
\begin{eqnarray}
    \Xi_f&=&\sum_{i=1}^N\log\frac{f(b_i-a_i)}{\epsilon}\nonumber\\
    &+&\sum_{i\ne j}\log\frac{f(|b_i-a_j|)}{\sqrt{f(|a_i-a_j|)f(|b_i-b_j|)}}
\end{eqnarray}
for positive functions $f$ on the interval $(0,L)$ (such as $\sigma$).
Note that we had to introduce a cutoff $\epsilon$ to regulate a divergence of the integral at the endpoints of the intervals; this is the usual divergence of the entropy in QFT. Substituting equation (\ref{intx}) into (\ref{renyiintx}), and changing variables from $\xi$ to $k$, we obtain
\begin{equation}\label{renyik}
    S_n=\frac{1}{\pi}\int_{\mathcal C} dk\left[2ik\Xi_\sigma-\ell\zeta_\nu(ik\ell)\right]r_n(k),
\end{equation}
where the integration path ${\mathcal C}$ is shown in figure \ref{fig2} and
\begin{eqnarray}
    r_n(k)=\frac{1}{1-n}\log\left[(1-e^{-2\pi k})^{-n}+(1-e^{2\pi k})^{-n}\right].
\end{eqnarray}
With
the further change of variables $k=q/2\pi\pm i/2$, and using that $\zeta_\nu$ is odd and commutes with complex conjugation, equation (\ref{renyik}) takes the form
\begin{equation}\label{renyidensity}
    S_n=\int_{-\infty}^\infty dq\, g(q)s_n(q)
\end{equation}
with
\begin{equation}\label{g}
    g(q)=\frac{1}{\pi^2}\left\{\Xi_\sigma-\ell\,{\text{Re}}\left[\zeta_\nu\left(\frac{\ell}{2}+\frac{i q\ell}{2\pi}\right)\right]\right\}
\end{equation}
and
\begin{equation}\label{sn}
    s_n(q)=\frac{1}{1-n}\log\left[(1+e^{-q})^{-n}+(1+e^q)^{-n}\right].
\end{equation}
Note that $s_n(q)$ is itself a Rényi entropy: it is the Rényi entropy of the probability distribution $p_0=1/(1+e^{-q})$, $p_1=1/(1+e^q)$ for a two-outcome experiment. 
The integral of this function is readily computed, 
\begin{equation}\label{integralsn}
    \int_{-\infty}^\infty dq\,s_n(q)=\frac{\pi^2}{6}\left(1+\frac{1}{n}\right),
\end{equation}
so we finally obtain
\begin{equation}\label{renyifinal}
    S_n=\frac{1}{6}\left(1+\frac{1}{n}\right)\Xi_\sigma+\Delta S_n,
\end{equation}
where
\begin{equation}\label{deltasn}
    \Delta S_n=-\frac{\ell}{\pi^2}\int_{-\infty}^\infty dq\,{\text{Re}}\left[\zeta_\nu\left(\frac{\ell}{2}+\frac{i q\ell}{2\pi}\right)\right]s_n(q).
\end{equation}
This equation can be rewritten in a slightly more compact (although perhaps less transparent) form by undoing our last change of variables,
\begin{equation}\label{deltasnk}
    \Delta S_n=-\frac{\ell}{\pi}\int_{\mathcal C} dk\,\zeta_\nu(ik\ell)r_n(k).
\end{equation}
Equation (\ref{renyifinal}), together with (\ref{deltasn}) or alternatively (\ref{deltasnk}), is the main result of this paper. 
It gives the Rényi entropies of a massless Dirac field on the torus, for an arbitrary spatial region and for arbitrary $n\in{\mathbb R}^+$. The form of this result is quite different from the series expansions of \cite{Herzog:2013py}, but figures \ref{fig3}, \ref{fig4} and \ref{fig5} show numerical evidence that both results are the same.
In these plots, the solid colored curves are computed with our expressions, and the dashed black curves are computed with equations (20), (21) and (43) of \cite{Herzog:2013py}. Below we will also show analytically that both results coincide in the case of integer $n$. Note that the first term in (\ref{renyifinal}) does not depend on the boundary conditions (i.e., on $\nu$), whereas the second term is insensitive to most details of the spatial region and only cares about its total length $\ell$.

The structure of our result for the Rényi entropy has a simple interpretation. To see this, let us go back to the general formula for the reduction of a Gaussian state to a spatial region $V$, equation (\ref{density}). Expanding $\psi$ in a basis of eigenfunctions of $K$ (which we will refer to as modes), the reduced density matrix factorizes as a product of single-mode density matrices,
\begin{equation}
    \rho=\prod_i \rho_i\qquad\rho_i=\frac{e^{-q_i a_{i}^\dagger a_i}}{1+e^{-q_i}},
\end{equation}
where $q_i$ is the eigenvalue of the $i$-th mode (we call it the modular energy of the mode) and $a_i$ is a fermionic annihilation operator, satisfying $\{a_i,a_j\}=0$ and $\{a_i,a_{j}^\dagger\}=\delta_{ij}$. In consequence, the Rényi entropy is a sum of single-mode Rényi entropies,
\begin{equation}
    S_n=\sum_i S_n(\rho_i)=\sum_i s_n(q_i),
\end{equation}
where $s_n$ is the function defined in (\ref{sn}). This formula can be rewritten in the form (\ref{renyidensity}), where $g$ is the density of modes, i.e., $g(q)dq$ is the number of modes with modular energies between $q$ and $q+dq$. Thus, equation (\ref{renyidensity}) is valid for generic Dirac fields in generic Gaussian states. Equation (\ref{g}) then gives the density of modes for the massless field on the torus.

\section{Particular cases}
\label{sect:4}

\subsection{Cylinder limits}

Let us see what the Rényi entropies look like when one of the periods of the torus is sent to infinity, so that the torus becomes a cylinder. Consider first the case $L\to\infty$. Using the form of $\sigma$ in this limit, equation (\ref{sigmainf}), we find
\begin{equation}\label{XiLinf}
    \Xi_\sigma=\Xi_v-\frac{1}{6}\left(\frac{\pi\ell}{\beta}\right)^2
\end{equation}
with
\begin{equation}
    v(x)=\frac{\beta}{\pi}\sinh\left(\frac{\pi x}{\beta}\right),
\end{equation}
whereas the form of $\sigma_\nu$, equation (\ref{sigmanuinf}), implies
\begin{equation}\label{zetanuLinf}
    \zeta_\nu(z)=-\frac{1}{3}\left(\frac{\pi}{\beta}\right)^2z.
\end{equation}
Substituting the above equations into (\ref{g}) we find that the density of modes is $g(q)=\Xi_v/\pi^2$, independent of $q$, so, by (\ref{renyidensity}) and (\ref{integralsn}), the Rényi entropy is
\begin{equation}\label{renyiLinf}
    S_n=\frac{1}{6}\left(1+\frac{1}{n}\right)\Xi_v.
\end{equation}
Not surprisingly, the dependence on the boundary conditions is lost in this limit. This result can also be recovered from the plane result of \cite{Casini:2005rm} by a suitable conformal transformation.

The limit $\beta\to\infty$ is richer, due to the fact that $\zeta_\nu$ has a slightly more complicated form, see equation (\ref{sigmanuinf}). Proceeding analogously to the previous case, we find
\begin{equation}\label{renyibetainf}
    S_n=\frac{1}{6}\left(1+\frac{1}{n}\right)\Xi_w+\delta_{\nu 0}\,\Delta_n(\ell),
\end{equation}
where
\begin{equation}
    w(x)=\frac{L}{\pi}\sin\left(\frac{\pi x}{L}\right)
\end{equation}
and
\begin{eqnarray}
    &&\Delta_n(\ell)=\int_{-\infty}^\infty dq\,h(\ell,q) s_n(q)\label{Deltan}\\
    &&h(\ell,q)=\frac{\ell}{\pi L}{\text{Re}}\left[\tan\left(\frac{\pi\ell}{2L}+\frac{iq\ell}{2L}\right)\right].
\end{eqnarray}
The integral (\ref{Deltan}) can be computed explicitly in the cases $\ell=0$ (where it vanishes) and $\ell=L$. To study the latter case, note that $h$ can be rewritten as
\begin{equation}
    h(\ell,q)=-\frac{\ell}{\pi L}{\text{Im}}\left\{\coth\left[\frac{q\ell}{2L}+\frac{i\pi(L-\ell)}{2L}\right]\right\}.
\end{equation}
Since $\coth(x+i\epsilon)=\coth x-i\pi\delta(x)$, where the principal part is understood in the first term, we have $h(L,q)=2\delta(q)$ and therefore
\begin{equation}\label{log4}
    \Delta_n(L)=2s_n(0)=\log 4.
\end{equation}
One can easily see (most easily in the case of a single interval) that $\Xi_w$ is invariant under the replacement of the region $V$ by its complement. Hence, for Neveu-Schwarz boundary conditions ($\nu=1$), equation (\ref{renyibetainf}) implies that the Rényi entropies of complementary regions coincide, as corresponds to a pure state (which in this case is the vacuum). This is shown in figure \ref{fig4}. The same is not true for Ramond boundary conditions ($\nu=0$), 
because $\Delta_n(L-\ell)\ne\Delta_n(\ell)$. The physical reason is that the massless field with Ramond boundary conditions has two constant modes (one for each chirality), which have zero energy, and hence the ground state is degenerate; it has degeneracy $4$ because each of the constant modes can be either empty or occupied. This means that the zero-temperature state is not a pure state, but a mixed state where each of these four vacua occurs with equal probability. The Rényi entropy of such a state is $\log 4$ for any value of $n$, which is precisely what we get from equations (\ref{renyibetainf}) and (\ref{log4}) when the spatial region is the whole circle, i.e., one interval of length $\ell=L-\epsilon$. The behavior of the Rényi entropy in the Ramond case is shown in figure \ref{fig5}. Equation (\ref{renyibetainf}) can also be recovered from the plane results of \cite{Casini:2005rm} by a conformal transformation in the Neveu-Schwarz case, but not in the Ramond case as far as we know.

\subsection{Entanglement entropy}

Taking the limit $n\to 1$ in our general formula for the Rényi entropy, equations (\ref{renyifinal}) and (\ref{deltasnk}), we obtain the entanglement entropy,
\begin{equation}
    S=\frac{1}{3}\Xi_\sigma+\Delta S,
\end{equation}
where 
\begin{equation}\label{deltas}
    \Delta S=-\frac{\ell}{\pi}\int_{\mathcal C} dk\,\zeta_\nu(ik\ell)r(k)
\end{equation}
and 
\begin{equation}\label{r(k)}
    r(k)=\frac{\log(1-e^{-2\pi k})}{1-e^{-2\pi k}}+\frac{\log(1-e^{2\pi k})}{1-e^{2\pi k}}.
\end{equation}
This result can be simplified. Indeed, since $\zeta_\nu$ only has poles at the points congruent to $L/2+\nu i\beta/2$, the integrand of (\ref{deltas})
is analytic on the strip bounded by ${\mathcal C}$ (see figure \ref{fig2}) except on the real axis, where $r$ has a cut.
Moreover, the integrand decays almost exponentially as ${\text{Re}}\,k\to\pm\infty$ ($\zeta_\nu$ grows, but only linearly). This means that both lines of ${\mathcal C}$ can be pushed towards the real axis, so that equation (\ref{deltas}) becomes
\begin{eqnarray}
    \Delta S&=&-\frac{\ell}{\pi}\int_{-\infty}^\infty dk\,\zeta_\nu(ik\ell)\left[r(k-i\epsilon)-r(k+i\epsilon)\right]\nonumber\\
    &=&2i\ell\left[\int_{-\infty}^0 dk\,\frac{\zeta_\nu(ik\ell)}{1-e^{-2\pi k}}-\int_{0}^\infty dk\,\frac{\zeta_\nu(ik\ell)}{1-e^{2\pi k}}\right]\nonumber\\
    &=&4i\ell\int_{0}^\infty dk\,\frac{\zeta_\nu(ik\ell)}{e^{2\pi k}-1},
\end{eqnarray}
where in the last step we have used that $\zeta_\nu$ is odd. Therefore, the entanglement entropy is
\begin{equation}\label{ee}
    S=\frac{1}{3}\Xi_\sigma+4i\ell\int_{0}^\infty dk\,\frac{\zeta_\nu(ik\ell)}{e^{2\pi k}-1}.
\end{equation}
This is the same result we had obtained previously in \cite{Blanco:2019cet} (up to a factor of $2$ because the result presented there corresponds to a single chirality).

\subsection{Integer order}

In the case $n\in{\mathbb Z}_{\ge 2}$, the integral in (\ref{deltasnk}) can be explicitly computed. To see this, let us introduce the function
\begin{equation}
    (L\sigma_\nu)(z)=\int_{0}^z dw\,\zeta_\nu(w)
\end{equation}
for $z$ on the strip $-L/2<{\text{Re}}\,z<L/2$, where the integration path is required to be contained entirely in that strip. Since $\zeta_\nu$ has no poles on the strip, $L\sigma_\nu$ is single-valued, i.e., at each point $z$ it is independent of the path chosen to go from $0$ to $z$. Moreover, it is analytic, $(L\sigma_\nu)'=\zeta_\nu$. Note that $(L\sigma_\nu)(z)$ is a logarithm of $\sigma_\nu(z)$, but it may not be the standard logarithm with imaginary part in $(-\pi,\pi)$ (the difference is an integer multiple of $2\pi i$, which may jump from point to point because the standard logarithm may have cuts). However, for real $x$ we do have
\begin{equation}\label{Lreal}
    (L\sigma_\nu)(x)=\int_{0}^x dy\,\zeta_\nu(y)=\log|\sigma_\nu(x)|,
\end{equation}
because $\zeta_\nu$ is real on the real line. Now, going back to (\ref{deltasnk}), note that the integrand decays almost exponentially as ${\text{Re}}\,k\to\pm\infty$, so we can close the contour of figure \ref{fig2} by adding to ${\mathcal C}$ the vertical sides of the rectangle,
\begin{equation}
    \Delta S_n=-\frac{\ell}{\pi}\ointctrclockwise dk\,\zeta_\nu(ik\ell)r_n(k).
\end{equation}
Integrating by parts we obtain
\begin{equation}
    \Delta S_n=\frac{1}{\pi i}\ointctrclockwise dk\,(L\sigma_\nu)(ik\ell)\,r_n'(k),
\end{equation}
and the explicit form of the above derivative is
\begin{equation}
    r_n'(k)=\frac{2\pi n}{1-n}\left[\frac{1}{1-e^{2\pi k}}-\frac{1}{1+(-e^{2\pi k})^n}\right].
\end{equation}
For generic $n$, this function has a cut on the real axis. However, for integer $n$ the cut disappears and $r_n'$ becomes meromorphic, so we can evaluate the integral by residues. The poles are at $k=im/n$ with $m=-(n-1)/2,-(n-1)/2+1,\dots,(n-1)/2$, so using (\ref{Lreal}) we obtain
\begin{equation}\label{integer}
    \Delta S_n=\frac{2}{1-n}\sum_{m=-\frac{n-1}{2}}^{\frac{n-1}{2}}\log|\sigma_\nu(m\ell/n)|.
\end{equation}
Together with (\ref{renyifinal}), this is the result obtained in \cite{Herzog:2013py,Azeyanagi:2007bj} for integer order. That result was expressed in terms of Jacobi theta functions; in appendix \ref{appendixB} we give the relation between the Weierstrass and the theta functions, and use it to verify the agreement a bit more explicitly.

We can use the result (\ref{integer}) to obtain the min-entropy $S_\infty$. Since the difference between two consecutive values of $m$ is $\Delta m=1$, we can rewrite (\ref{integer}) as
\begin{equation}
    \Delta S_n=\frac{2n}{1-n}\sum_{m=-\frac{n-1}{2}}^{\frac{n-1}{2}}\Delta(m/n)\log|\sigma_\nu(m\ell/n)|.
\end{equation}
In the limit $n\to\infty$ this sum becomes an integral, 
\begin{equation}
    \Delta S_\infty=-2\int_{-1/2}^{1/2}dx\,\log|\sigma_\nu(x\ell)|,
\end{equation}
so the min-entropy is
\begin{equation}
    S_\infty=\frac{1}{6}\Xi_\sigma-2\int_{-1/2}^{1/2}dx\,\log|\sigma_\nu(x\ell)|.
\end{equation}
This gives the largest eigenvalue of the reduced density matrix. Of course, this eigenvalue is very small, and vanishes when the cutoff is removed. This is not special of the case we are considering: the reduced density matrix is not well-defined in QFT.

\section{Rényi mutual information}
\label{sect:5}

\begin{figure}
    \centering
    \includegraphics[width=0.34\textwidth]{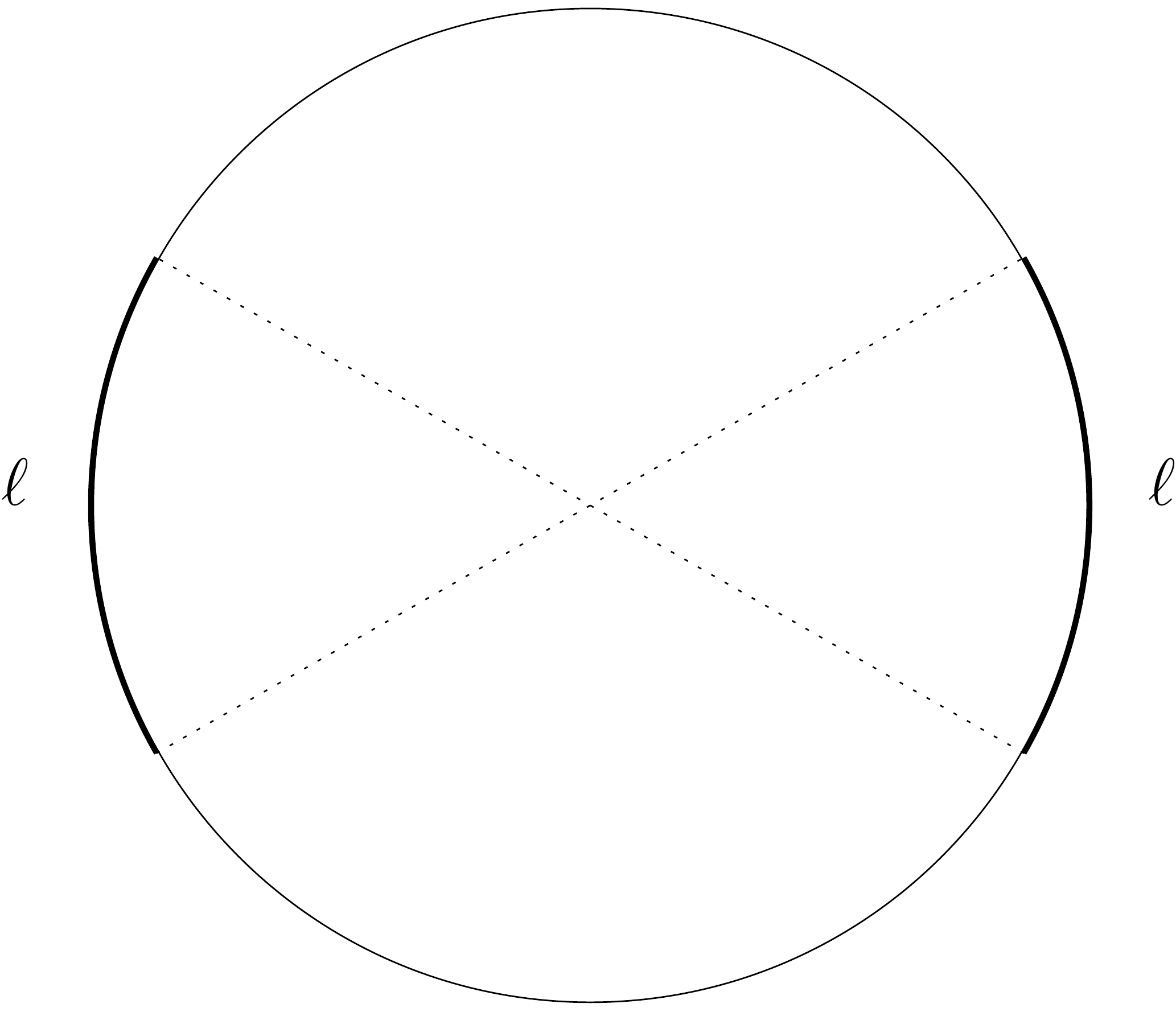}
    \caption{Pair of intervals for which we compute the RMI.}
    \label{fig6}
\end{figure}

With our result for the Rényi entropies, it is straightforward to compute any RMI. For the simple configuration of figure \ref{fig6}
(two intervals of length $\ell$ arranged symmetrically), the RMIs have the form
\begin{equation}\label{mutual}
    I_n=\frac{1}{6}\left(1+\frac{1}{n}\right)F+\Delta I_n,
\end{equation}
where 
\begin{equation}
    F(\ell)=\log\left[\frac{\sigma^2(L/2)}{\sigma(L/2-\ell)\sigma(L/2+\ell)}\right]
\end{equation}
and $\Delta I_n$ is the contribution from the second term in (\ref{renyifinal}). We display it explicitly for two extreme values of $n$,
\begin{eqnarray}
    &&\Delta I(\ell)=8i\int_{0}^\infty dx\,\frac{\zeta_\nu(ix)-\zeta_\nu(2ix)}{e^{2\pi x/\ell}-1}\\
    &&\Delta I_\infty(\ell)=-2\int_{-1/2}^{1/2}dx\,\log\left|\frac{\sigma_{\nu}^2(x\ell)}{\sigma_\nu(2x\ell)}\right|
\end{eqnarray}
(the absence of subscript means $n=1$).
Note that the dependence on the cutoff has dropped out from (\ref{mutual}), as expected: the RMIs are finite.

Let us check that the mutual information ($n=1$) is positive and monotonic, as it should be.  
Since it vanishes at $\ell=0$, as is clear from the above equations, 
it suffices to show that it is monotonic, $I'(\ell)\ge 0$. We have
\begin{equation}
    F'(\ell)=\zeta(L/2-\ell)-\zeta(L/2+\ell)
\end{equation}
and
\begin{equation}
    \Delta I'(\ell)=\frac{4\pi i}{\ell^2}\int_{0}^\infty dx\,\frac{x\left[\zeta_\nu(ix)-\zeta_\nu(2ix)\right]}{\sinh^2(\pi x/\ell)}.
\end{equation}
The above differences of zeta functions can be bounded using the properties of the Weierstrass elliptic function
\begin{equation}
    \wp=-\zeta',
\end{equation}
which is periodic, $\wp(z+P_i)=\wp(z)$, hence the name (elliptic means meromorphic and periodic). We also define
\begin{equation}
    \wp_\nu(z)=-\zeta_\nu'(z)=\wp(z+L/2+\nu i\beta/2).
\end{equation}
As shown in appendix \ref{appendixC} (see the upper-left panel of figure \ref{fig8}), for real $x$ we have $\wp(x)\ge\wp(L/2)$. Integrating this inequality over the interval $(L/2-\ell,L/2+\ell)$ we obtain
\begin{equation}\label{ineq1}
    F'(\ell)\ge 2\ell\wp(L/2).
\end{equation}
On the other hand, it is clear from the lower-right panel of figure \ref{fig8} that $\wp_\nu(iy)\le\wp(L/2)$ for real $y$. Integrating this inequality in $y$ from $x$ to $2x$, with $x$ positive, we find
\begin{equation}
    i\left[\zeta_\nu(ix)-\zeta_\nu(2ix)\right]\ge -\wp(L/2)x,
\end{equation}
so
\begin{eqnarray}\label{ineq2}
    \Delta I'(\ell)&\ge&-\frac{4\pi }{\ell^2}\wp(L/2)\int_{0}^\infty dx\,\frac{x^2}{\sinh^2(\pi x/\ell)}\nonumber\\
    &=&-\frac{2\ell}{3}\wp(L/2).
\end{eqnarray}
Substituting the inequalities (\ref{ineq1}) and (\ref{ineq2}) into (\ref{mutual}) with $n=1$ we find that $I'(\ell)\ge 0$, as we wanted to show. Hence, the mutual information is positive and monotonic, as it should be.

\begin{figure}
    \centering
    \includegraphics[width=0.4\textwidth]{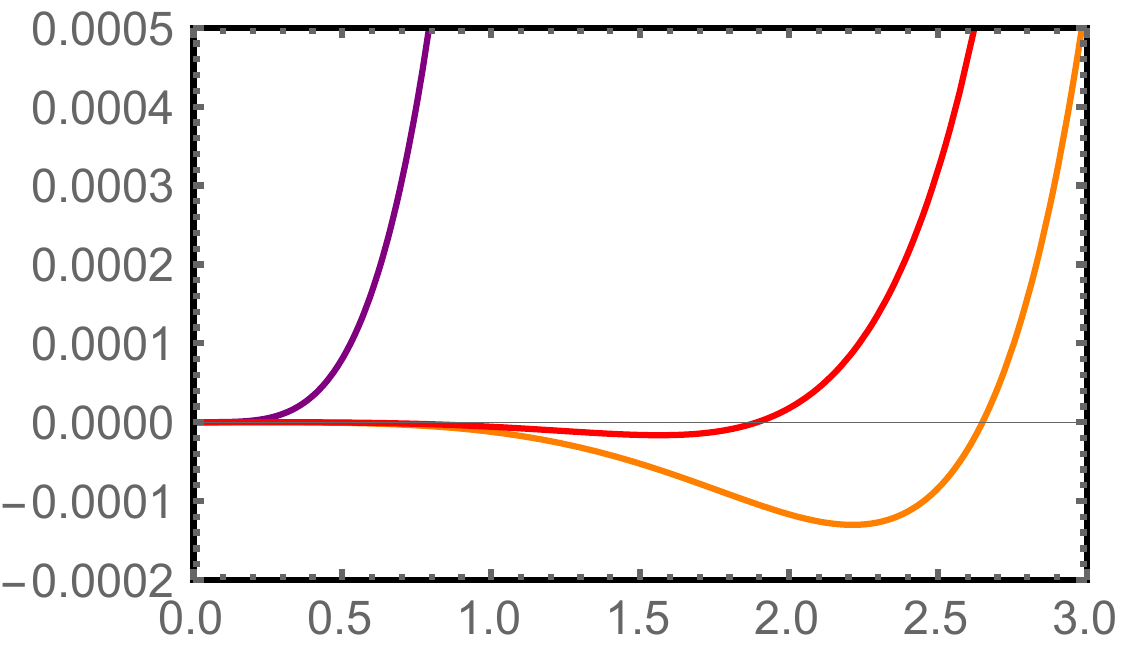}
    \caption{Ramond RMIs for the configuration of figure \ref{fig6} as a function of $\ell$, for $n=1$ (top curve), $n=10$ (middle) and $n=\infty$ (bottom), with $L=10$, $\beta=50$. The top curve is positive and monotonic, but the other two are not.}
    \label{fig7}
\end{figure}

Unlike the mutual information, the RMIs with $n\ne 1$ are not expected to be positive or monotonic in general. Let us study the behavior of $I_\infty$. Expanding to fourth order around $\ell=0$ we find
\begin{equation}
    F(\ell)=\wp(L/2)\ell^2+\frac{1}{12}\wp''(L/2)\ell^4+{\mathcal O}(\ell^6)
\end{equation}
and
\begin{equation}
    \Delta I_\infty(\ell)=-\frac{1}{6}\wp_\nu(0)\ell^2-\frac{7}{480}\wp_\nu''(0)\ell^4+{\mathcal O}(\ell^6),
\end{equation}
which after substitution into (\ref{mutual}) yields
\begin{eqnarray}
   \!\!\!\!\!\!\!\!\! I_\infty(\ell)&=&\frac{1}{6}\left[\wp(L/2)-\wp_\nu(0)\right]\ell^2\nonumber\\
    &+&\frac{1}{12}\left[\frac{1}{6}\wp''(L/2)-\frac{7}{40}\wp_\nu''(0)\right]\ell^4+{\mathcal O}(\ell^6).
\end{eqnarray}
Now, the lower-right panel of figure \ref{fig8} makes it clear that $\wp_1(0)<\wp(L/2)$, so in the Neveu-Schwarz case, $\nu=1$, the first term above is positive and hence $I_\infty$ is positive and monotonic, at least near $\ell=0$. In contrast, in the Ramond case, $\nu=0$, the first term above cancels and we have
\begin{equation}
    I_\infty(\ell)=-\frac{1}{1440}\wp''(L/2)\ell^4+{\mathcal O}(\ell^6).
\end{equation}
Since $\wp''(L/2)>0$, it follows
that $I_\infty$ is non-positive and non-monotonic. This is shown in figure \ref{fig7}, where we also show how positivity and monotonicity are recovered as $n\to 1$.
Note that the effect is small (the temperature in figure \ref{fig7} is chosen so as to make $\wp''(L/2)$ as large as possible), but nevertheless it is still there.

\section{Discussion}
\label{sect:6}

In this paper we computed the Rényi entropies of the massless Dirac field on the torus for arbitrary spatial regions. These quantities had already been computed in \cite{Herzog:2013py,Azeyanagi:2007bj}, but by a different method, the replica trick. Here we used the resolvent method, i.e., we expressed the entropies in terms of the resolvent of the two-point function (thought of as an operator) and used the explicit form of that resolvent obtained recently in \cite{Blanco:2019xwi,Blanco:2019cet}. Our results are in perfect agreement with those of \cite{Herzog:2013py,Azeyanagi:2007bj} but have a different form (they involve an integral instead of an infinite series). Thus, they serve as a cross-check and also provide an alternative representation for the Rényi entropies, which may be useful for numerical applications. 

We also computed the RMIs in simple cases, and found that, for appropriate choices of the parameters, they are non-positive and non-monotonic. This behavior is expected, but we are not aware of previous examples in the context of QFT. The simplest known Rényi entropies (those of the massless Dirac field on the plane and their conformal transformations to the cylinders) do not have this behavior. The violation of positivity and monotonicity we find is very small; it would be interesting to know if there is a general reason behind this or it is just a peculiarity of the case we are considering.

\appendix

\section{Relation between Weierstrass and Jacobi theta functions}\label{appendixB}

Following the conventions of \cite{francesco1997conformal} for the Jacobi theta functions $\theta_i$ ($i=1,2,3,4$), we have (see theorem 2, chapter V of \cite{chandrasekharan2012elliptic})
\begin{equation}
    \sigma(z)=L\frac{\theta_1(z/L)}{\theta_1'(0)}\exp\left(\eta z^2\right),
\end{equation}
where $\eta=\zeta(L/2)/L$ and we are omitting the depencence of $\theta_i$ on the modular parameter $\tau=i\beta/L$.
This implies
\begin{equation}
   \sigma_\nu(z)=\frac{\theta_{\nu+2}(z/L)}{\theta_{\nu+2}(0)}\exp\left(\eta z^2\right).
\end{equation}
These equations can be used to rewrite all the equations of this paper involving Weierstrass functions in terms of theta functions. In particular, setting $L=1$ we have
\begin{equation}
    \Xi_\sigma=\Xi_{\theta_1/\theta_1'(0)}+\eta\ell^2
\end{equation}
and, for $n\in{\mathbb Z}$,
\begin{eqnarray}
    &&\sum_{m=-\frac{n-1}{2}}^{\frac{n-1}{2}}\log\left|\sigma_\nu\left(\frac{m\ell}{n}\right)\right|\ \,=\sum_{m=-\frac{n-1}{2}}^{\frac{n-1}{2}}\log\left|\frac{\theta_{\nu+2}\left(\frac{m\ell}{n}\right)}{\theta_{\nu+2}(0)}\right|\nonumber\\
    &&\ \ \ \ \ \ \ \ \ \ \ \ \ \ \ \ \ \ \ \ \ \ \ \ \  +\ \frac{1}{12}(n-1)\left(1+\frac{1}{n}\right)\eta\ell^2.
\end{eqnarray}
Substituting the above two equations into (\ref{renyifinal}) and (\ref{integer}), we find that our result for integer $n$ coincides exactly with equations (20)-(22) of \cite{Herzog:2013py}.

\section{Weierstrass elliptic function}
\label{appendixC}

The properties of the Weierstrass elliptic function $\wp$ follow easily from those of $\zeta$, together with general properties of elliptic functions. It has a pole at the origin, around which it behaves as $\wp(z)=1/z^2+\dots$, and no other poles except for the congruent points. It is even, which, together with periodicity, implies the additional parity $\wp(\omega_i-z)=\wp(\omega_i+z)$ with respect to the points $\omega_1=L/2$, $\omega_2=i\beta/2$ and $\omega_3=L/2+i\beta/2$. Since these points are not poles (because they are not congruent to the origin), it follows that they are stationary points of $\wp$ or, in other words, zeros of $\wp'$. A basic property of elliptic functions is that, within a cell (i.e., a rectangle of sides $L$ and $i\beta$), the number of poles is equal to the number of zeros, all weighted by their multiplicity. Clearly, $\wp'$ is an elliptic function with a triple pole at the origin and no other poles except for the congruent points, so it must have exactly $3$ zeros within a cell, which implies that $\omega_1$, $\omega_2$ and $\omega_3$ are the only stationary points of $\wp$ except for the congruent points. Finally, $\wp$ commutes with complex conjugation, which, together with its parity properties, implies that it is real on the real axis and its translations by multiples of $i\beta/2$, and also on the imaginary axis and its translations by multiples of $L/2$.

\begin{figure}
    \centering
    \includegraphics[width=0.483\textwidth]{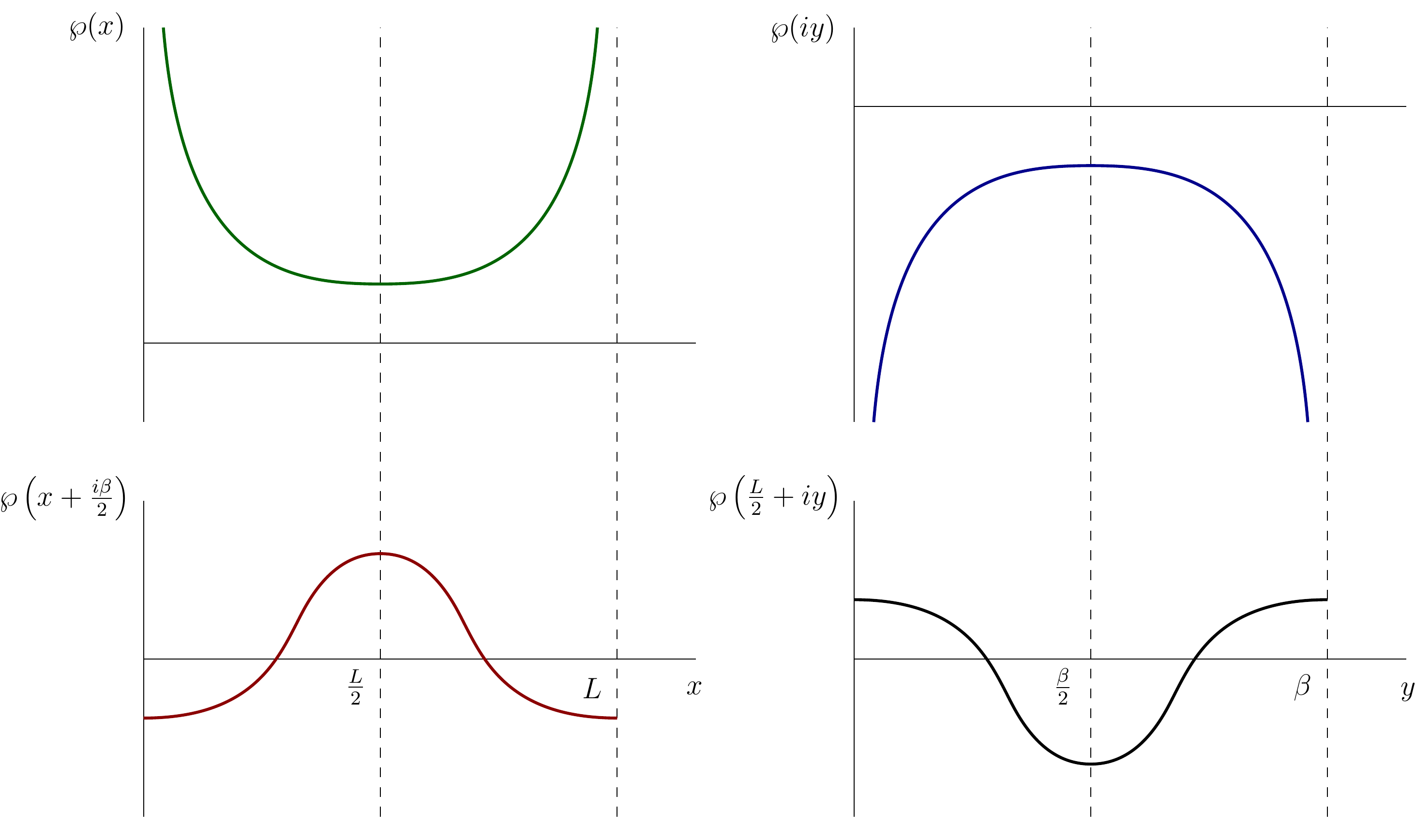}
    \caption{Qualitative behavior of the Weierstrass elliptic function on the lines where it is real.}
    \label{fig8}
\end{figure}

Let us see what $\wp$ looks like on the lines where it is real. It suffices to look at the intervals $(0,L)$, $i(0,\beta)$,  $(0,L)+i\beta/2$ and $L/2+i(0,\beta)$; the behavior everywhere else is determined by periodicity. Note that (taking seriously that the intervals are open), $\wp$ has no poles on any of the intervals, and its only stationary point within each interval is the middle point, around which it is symmetric. It follows that the middle point is either an absolute maximum or an absolute minimum within the interval. Now, for $x\in{\mathbb R}$ we have $\wp(x)\to+\infty$ and $\wp(ix)\to-\infty$ as $x\to 0$. Therefore, the middle point of $(0,L)$ is a minimum, and that of $i(0,\beta)$ is a maximum. This means $\wp''(L/2),\wp''(i\beta/2)\ge 0$, which in turn implies that the middle point of the interval $(0,L)+i\beta/2$ is a maximum and that of $L/2+i(0,\beta)$ is a minimum. This discussion leads to the qualitative plot of figure \ref{fig8}.

{\sl Acknowledgements.---} This work has been partially supported by CONICET, UBA and through the grant UBACyT 20020190200162BA. DB acknowledges support from the ANPCyT through the grant PICT-2018-03593.

\bibliography{references}
\bibliographystyle{ieeetr}

\end{document}